\documentclass[12pt]{article}
\usepackage{epsfig}
\usepackage{amssymb}
\usepackage{amsmath}
\usepackage{amsfonts}

\newcommand{\R}{\mathbb{R}}
\newcommand{\C}{\mathbb{C}}
\newcommand{\Z}{\mathbb{Z}}

\newcommand{\be}{\begin{equation}}
\newcommand{\ee}{\end{equation}}
\newcommand{\bea}{\begin{eqnarray}}
\newcommand{\eea}{\end{eqnarray}}
\newcommand{\nn}{\nonumber}
\newcommand{\kt}{\rangle}
\newcommand{\br}{\langle}

\newcommand{\ed}{\end{document}}

\newcommand{\np}{\newpage}
\newcommand{\dl}{[\![}
\newcommand{\dr}{]\!]}

\begin{document}

\title{Metric Operator in Pseudo-Hermitian Quantum Mechanics
and the Imaginary \\Cubic Potential}

\author{\\
Ali Mostafazadeh
\\
\\
Department of Mathematics, Ko\c{c} University,\\
34450 Sariyer, Istanbul, Turkey\\ amostafazadeh@ku.edu.tr}
\date{ }
\maketitle

\begin{abstract}
We present a systematic perturbative construction of the most
general metric operator (and positive-definite inner product) for
quasi-Hermitian Hamiltonians of the standard form,
$H=\frac{1}{2}\,p^2+v(x)$, in one dimension. We show that this
problem is equivalent to solving an infinite system of iteratively
decoupled hyperbolic partial differential equations in
$(1+1)$-dimensions. For the case that $v(x)$ is purely imaginary,
the latter have the form of a nonhomogeneous wave equation which
admits an exact solution. We apply our general method to obtain
the most general metric operator for the imaginary cubic
potential, $v(x)=i\epsilon\,x^3$. This reveals an infinite class
of previously unknown ${\cal CPT}$- as well as non-${\cal
CPT}$-inner products. We compute the physical observables of the
corresponding unitary quantum system and determine the underlying
classical system. Our results for the imaginary cubic potential
show that, unlike the quantum system, the corresponding classical
system is not sensitive to the choice of the metric operator. As
another application of our method we give a complete
characterization of the pseudo-Hermitian canonical quantization of
a free particle moving in $\R$ that is consistent with the usual
choice for the quantum Hamiltonian. Finally we discuss subtleties
involved with higher dimensions and systems having a fixed length
scale.

\vspace{5mm}

\vspace{2mm}


\end{abstract}




\np

\section{Introduction}

In Quantum Mechanics the unitarity of the dynamics dictates that
the Hamiltonian operator must be  Hermitian (self-adjoint).
Therefore a non-Hermitian operator $H$ cannot serve as the
Hamiltonian for a unitary quantum system unless the Hilbert space
is endowed with a different inner-product which renders $H$
Hermitian. Under rather general conditions, one can show that a
necessary and sufficient condition for the existence of such an
inner product is the reality of the spectrum of $H$, \cite{p2p3}.
The most general inner product that makes $H$ Hermitian has the
form
    \be
    \br\cdot,\cdot\kt_{\eta_+}:=\br\cdot|\eta_+|\cdot\kt,
    \label{inn}
    \ee
where $\br\cdot|\cdot\kt$ is the defining inner product of the
original (reference) Hilbert space ${\cal H}$, $\eta_+:{\cal
H}\to{\cal H}$ is a positive-definite operator satisfying the
pseudo-Hermiticity condition \cite{p1}
    \be
    H^\dagger=\eta_+ H\eta_+^{-1},
    \label{ph}
    \ee
and $H^\dagger:{\cal H}\to{\cal H}$ is the adjoint of
$H$.\footnote{For any densely defined $H$, $H^\dagger$ is the
unique operator satisfying $\br\cdot|H^\dagger\,\cdot\kt=\br
H\,\cdot|\cdot\kt$, \cite{simon-reed}.}

The operator $\eta_+$, which is sometimes referred to as the
\emph{metric operator}, is the basic ingredient of a quantum
theory based on the Hamiltonian $H$. It determines the inner
product (\ref{inn}) of the physical Hilbert space ${\cal H}_{\rm
phys}$ and specifies the observables $O$ of the theory,
\cite{critique,jpa04b,jpa05b,p66}. Furthermore, it provides the
necessary means to obtain an equivalent description of the system
using a Hermitian Hamiltonian $h$ acting in ${\cal H}$. This in
turn allows for the determination of the underlying classical
system, \cite{jpa04b,jpa05b}.

An important property of the metric operator, is that it is not
unique \cite{p4,jmp2003,geyer,jones}. For example, the ${\cal
CPT}$-inner product proposed in \cite{bbj} can be conveniently
expressed in the form (\ref{inn}) for a particular choice of
$\eta_+$, \cite{jmp2003,jpa05a}. As such, the ${\cal CPT}$-inner
product is not the only allowed choice for the inner product of
the physical Hilbert space; it is merely an example of an infinite
set of possible choices.

In \cite{jpa04b,jpa05b,p66} we have constructed specific metric
operators for different ${\cal PT}$-symmetric systems and examined
the physical consequences and the classical limit of the quantum
systems they define. In the present article we intend to address
the issue of the non-uniqueness of the metric operator. In
particular, we will offer a perturbative construction of the most
general metric operator for the standard Hamiltonians in one
dimension, and explore the consequences of adopting different
metric operators for a given Hamiltonian operator.

Our method which is based on a variation of the approach of
\cite{bbj-prd} has the following two advantages. (\emph{i})~It
does not involve using a particular ansatz for each choice of the
potential. Instead, it employs the well-established properties of
$(1+1)$-dimensional hyperbolic partial differential equations.
This makes it applicable for all potentials. (\emph{ii})~It is not
restricted to producing the ${\cal CPT}$-inner products; it yields
the most general inner product that is capable of restoring the
unitarity. In fact, as we will see for the particular example of
the imaginary cubic potential, it reveals a new set of ${\cal
CPT}$-inner products that has escaped the analysis of
\cite{bbj-prd}.

\section{Perturbative Construction of the Metric Operator}

Consider a Hamiltonian of the form
    \be
    H=H_0+\epsilon H_1,
    \label{H=HH}
    \ee
where $\epsilon$ is a real perturbation parameter and $H_0$ and
$H_1$ are respectively Hermitian and anti-Hermitian
$\epsilon$-independent operators. Suppose that for sufficiently
small values of $\epsilon$ the Hamiltonian $H$ has a real spectrum
and a complete set of eigenvectors, so that a positive-definite
metric operator $\eta_+$ exists \cite{p2p3}. Such an operator has
a well-defined Hermitian logarithm. We shall let $Q:=-\ln\eta_+$,
or alternatively
    \be
    \eta_+=e^{-Q},
    \label{log}
    \ee
and employ (\ref{ph}) to obtain a perturbative expansion for $Q$.

The condition that for $\epsilon=0$, the inner product
$\br\cdot,\cdot\kt_{\eta_+}$ of (\ref{inn}) reduces to
$\br\cdot|\cdot\kt$, i.e., as $\epsilon\to 0$ we have $\eta_+\to
1$, justifies the following perturbation expansion for $Q$.
    \be
    Q=\sum_{j=1}^\infty Q_j\epsilon^j,
    \label{expand}
    \ee
where $Q_j$ are $\epsilon$-independent Hermitian operators. A
convenient tool for calculating the operators $Q_j$ is the
Baker-Campbell-Hausdorff formula,
    \be
    e^{-Q}H\,e^{Q}=H+\sum_{k=1}^\infty \frac{1}{k!}[H,Q]_{_k},
    \label{bch}
    \ee
where $[H,Q]_{_1}:=[H,Q]$ and $[H,Q]_{_{k+1}}:=[[H,Q]_{_k},Q]$ for
all $k\in\Z^+:=\{1,2,3,\cdots\}$. As we show in Appendix~A, we can
use (\ref{bch}) to obtain the following infinite set of operator
equations for $Q_j$.
    \be
    \left[H_0,Q_j\right]= R_j,
    \label{e}
    \ee
where $j\in\Z^+$ is arbitrary and
   \bea
    R_j&:=&\left\{\begin{array}{cc}
    -2H_1~~~~
    &{\rm for} \quad j=1\\
    \sum_{k=2}^jq_k\,Z_{kj}&{\rm for} \quad j\geq 2,
    \end{array}\right.
    \label{R-combi}\\
    q_{k}&:=&\sum_{m=1}^k\sum_{n=1}^m
    \frac{(-1)^n n^k}{k!2^{m-1}}
    \mbox{\small$\left(\!\!\begin{array}{c} m\\n\end{array}\!\!
    \right)$},
    \label{q-mk}\\
    Z_{kj}&:=&\hspace{-.5cm}
    \sum_{
    \mbox{\scriptsize$\begin{array}{c}
    s_1,s_2,\cdots,s_k\in\Z^+\\
    s_1+\cdots+s_k=j
    \end{array}$}}[[\cdots[H_0,Q_{s_1}],Q_{s_2}],\cdots,],Q_{s_k}].
    \label{chi=}
    \eea
It is not difficult to see that the anti-Hermitian operators $R_j$
are uniquely determined by $H_1$, $Q_1$, $Q_2,\cdots,Q_{j-1}$.
This shows that (\ref{e}) may in principle be solved iteratively.
For example setting $j=1,2,3,4$ and $5$, we respectively find
    {\small
    \bea
    \left[H_0,Q_1\right]&=&-2H_1,
    \label{e1}\\
    \left[H_0,Q_2\right]&=& 0,
    \label{e2}\\
    \left[H_0,Q_3\right]&=&\frac{1}{12}[H_0,Q_1]_{_3}=
    -\frac{1}{6}[H_1,Q_1]_{_2},
    \label{e3}\\
    \left[H_0,Q_4\right]&=&\frac{1}{12}\,\left(
    [[H_0,Q_1]_{_2},Q_2]+[[[H_0,Q_1],Q_2],Q_1]+[[H_0,Q_2],Q_1]_{_2}
    \right)\nn\\
    &=&-\frac{1}{6}\left([[H_1,Q_1],Q_2]+[[H_1,Q_2],Q_1]\right),
    \label{e4}\\
    \left[H_0,Q_5\right]&=&-\frac{1}{120}\,[H_0,Q_1]_{_5}+
    \frac{1}{12}\left(
    [[H_0,Q_1],Q_2]_{_2}+[[H_0,Q_2],Q_1]_{_2}+[[H_0,Q_2]_{_2},Q_1]+
    \right.\nn\\
    &&\left.
    [[H_0,Q_1]_{_2},Q_3]+[[[H_0,Q_1],Q_3],Q_1]+[[H_0,Q_3],Q_1]_{_2}
    \right)\nn\\
    &=&\frac{1}{360}\,[H_1,Q_1]_{_4}-\frac{1}{6}\,
    \left([H_1,Q_2]_{_2}+[[H_1,Q_1],Q_3]+[[H_1,Q_3],Q_1]\right),
    \label{e5}
    \eea}
where we have made use of the fact that $q_k$ vanishes for even
values of $k$.

Comparing (\ref{e1}) -- (\ref{e5}) with operator equations
obtained in \cite{bbj-prd} for $Q_j$, we see that the choice of
the ${\cal CPT}$-inner product corresponds to taking
$Q_2=Q_4=\cdots=Q_{2j}=\cdots=0$. Furthermore this choice
restricts $Q_{2j+1}$ to the solutions of (\ref{e}) that are even
in $x$ and odd in $p$. We will refrain from imposing such
restrictions and try to obtain the most general solution of
(\ref{e}).

Setting $H_0=\frac{1}{2m}\,p^2+v_0(x)$, where $v_0$ is a
real-valued function, adopting units in which $m=\hbar=1$, and
evaluating the matrix-elements of both sides of (\ref{e}) in the
$x$-representation, we find
    \be
    [-\partial_x^2+\partial_y^2+2v_0(x)-2v_0(y)]\br x|Q_j|y\kt=
    2\br x|R_j|y\kt.
    \label{pde}
    \ee
This is a nonhomogeneous hyperbolic partial differential equation
in $(1+1)$-dimensions. In particular, for the cases that $v_0=0$,
i.e., $H_0=\frac{1}{2}\,p^2$, it reduces to the
$(1+1)$-dimensional nonhomogeneous wave equation,
    \be
    [-\partial_x^2+\partial_y^2]\br x|Q_j|y\kt=
    2\br x|R_j|y\kt,
    \label{weq}
    \ee
whose general closed-form solution is well-known \cite{wave-eqn}.

In summary, we have reduced the problem of determining $Q_j$ to
the solution of a set of iteratively decoupled and well-known
partial differential equations. In fact, because these equations
only differ in the form of their nonhomogeneity, we can express
their general solution as
    \be
    \br x|Q_j|y\kt= f_j(x,y)+{\cal Q}_j(x,y),
    \label{generic}
    \ee
where $f_j$ is the general solution of the homogeneous equation
    \be
    [-\partial_x^2+\partial_y^2+2v_0(x)-2v_0(y)]f_j(x,y)=0,
    \label{pde-0}
    \ee
and ${\cal Q}_j$ is a particular solution of (\ref{pde}). The only
restriction on $f_j$ and ${\cal Q}_j$ is that they should be
selected in such a way that $\br x|Q_j|y\kt^*=\br y|Q_j|x\kt$,
i.e., yield a Hermitian $Q_j$. We can adopt, without loss of
generality, a particular solution satisfying the Hermiticity
condition ${\cal Q}_j(x,y)={\cal Q}_j(y,x)^*$. In this case
$f_j(x,y)=f_j(y,x)^*$ and as a result the general form of $f_j$ is
independent of $j$.\footnote{We will however continue using the
label $j$, because for different $j$ we can adopt different
homogeneous solutions.}

For the case that $H_0=\frac{1}{2}\,p^2$, $f_j$ is given by
d'Alemert's formula \cite{wave-eqn}:
    \be
    f_j(x,y)=\varphi_j(x-y)+\chi_j(x+y),
    \label{fg=}
    \ee
where $\varphi_j,\chi_j:\R\to\C$ are arbitrary twice
differentiable functions.\footnote{The Hermiticity condition:
$f_j(x,y)=f_j(y,x)^*$ imposes the following restrictions on
$\varphi_j$ and $\chi_j$: $\Re[\varphi_j(-x)]=\Re[\varphi_j(x)]$,
$\Im[\varphi_j(-x)]=-\Im[\varphi_j(x)]-2r_j$, $
\Im[\chi_j(x)]=r_j$, where $\Re$ and $\Im$ stand for real and
imaginary parts of their argument respectively and $r_j$ is an
arbitrary real constant.\label{condi}} Because, according to
(\ref{e2}), $R_2=0$, we can set ${\cal Q}_2(x,y)=0$ and use
(\ref{generic}) and (\ref{fg=}), to obtain
    \be
    \br x|Q_2|y\kt= \varphi_2(x-y)+\chi_2(x+y).
    \label{Q2=}
    \ee
We can easily obtain the explicit form of the operator $Q_2$ using
(\ref{Q2=}). The result is
$Q_2=\tilde\varphi_2(p)+\tilde\chi_2(p)\,{\cal P}$, where a tilde
stands for the Fourier transform: $\tilde
F(p):=\int_{-\infty}^\infty e^{-ipx}F(x)dx$, and ${\cal P}$ is the
parity operator. The same reasoning shows that, whenever
$H_0=\frac{1}{2}\,p^2$,
    \be
    Q_j={\cal Q}_j+F_j+K_j\,{\cal P},
    \label{Qj=}
    \ee
where $j\in\Z^+$ is arbitrary, ${\cal Q}_j$ is a particular
solution of the operator equation (\ref{e}), and
$F_j=\tilde\varphi_j(p)$ and $K_j=\tilde\chi_j(p)$ are any
functions of $p$ that together with ${\cal Q}_j$ make $Q_j$
Hermitian.\footnote{They must also be piece-wise analytic and
possess (inverse) Fourier transforms.} In particular, if we select
${\cal Q}_j$ to be Hermitian, $F_j+K_j\,{\cal P}$ will be
Hermitian.

It turns out that the condition that $F_j+K_j\,{\cal P}$ be
Hermitian, which can be stated as
    \be
    {\cal P}\,K_j^\dagger-K_j{\cal P}=F_j-F_j^\dagger,
    \label{hermiticity}
    \ee
is equivalent to the requirement that $F_j$ and $K_j$ be
respectively Hermitian and ${\cal P}$-pseudo-Hermitian \cite{p1},
i.e.,
    \be
    F_j^\dagger=F_j,~~~~~~~K_j^\dagger=
    {\cal P}K_j{\cal P}^{-1}={\cal P}K_j{\cal P}.
    \label{p-ph}
    \ee
Therefore, as functions of $p$, $F_j$ is real-valued while $K_j$
is ${\cal PT}$-invariant.\footnote{This means that for any real
variable $p_c$, $F_j(p_c)\in\R$ and $K_j(p_c)^*=K_j(-p_c)$.} One
way of seeing this is to use the identities $[p,{\cal P}]=2p{\cal
P}$ and $[p,K_j]=0$ to show that
    \be
    [p,{\cal P}\,K_j^\dagger-K_j{\cal P}]=
    2p({\cal P}K_j^\dagger-K_j{\cal P}).
    \label{e31}
    \ee
In view of (\ref{hermiticity}) and $[p,F_j]=0$, the left-hand side
of (\ref{e31}) vanishes identically whereas its right-hand side
equals $2p(F_j-F_j^\dagger)$. Hence, $F_j-F_j^\dagger=0$, which
together with (\ref{hermiticity}) imply (\ref{p-ph}).

For the case that $H_0$ involves a non-vanishing potential $v_0$,
equations~(\ref{pde}) and (\ref{pde-0}) do not admit a closed form
exact solution. In fact, one can prove that the factorization
technique that is at the heart of the derivation of d'Alemert's
formula (\ref{fg=}) does not apply to (\ref{pde-0}). The spectral
method that is also originally developed to solve wave (and heat)
equations admit a generalization to both (\ref{pde}) and
(\ref{pde-0}). Yet it does not produce generic closed form
expressions for the most general solutions of (\ref{pde}) and
(\ref{pde-0}).

In the remainder of this paper we will only be concerned with the
case $v_0=0$, i.e., $H_0=\frac{1}{2}\,p^2$. In this case we can
use the following prescription to compute ${\cal Q}_j$.
    \begin{itemize}
    \item[({\em i})] Fourier transform both sides of
(\ref{weq}) over $y$ to yield the ordinary differential equation:
    \be
    (\partial_x^2+p^2)\br x|Q_j|p\kt=-2\br x|R_j|p\kt,
    \label{ode}
    \ee
where $p$ is treated as a free real parameter. This equation
admits solutions of the form
    \be
    -2\int_{x_0}^x dx'\,G_p(x,x')\br x'|R_j|p\kt
    \label{integral}
    \ee
where $G_p(x,x')=\sin[p(x-x')]/p$ is the Green's function for
$\partial_x^2+p^2$;
    \item[({\em ii})] Evaluate the integral in (\ref{integral})
by making a simple choice for $x_0$ and identify the resulting
particular solution of (\ref{ode}) with $\br x|{\cal Q}_j|p\kt$.
    \item[({\em iii})] Take the inverse Fourier transform of
$\br x|{\cal Q}_j|p\kt$ over $p$ and denote the result by $\br
x|{\cal Q}_j|y\kt$. If $\br x|{\cal Q}_j|y\kt^*=\br y|{\cal
Q}_j|x\kt$ proceed to ({\em iv}). If not, add the homogeneous
solutions $u_\pm(p)\,e^{\pm ipx}$ to $\br x|{\cal Q}_j|p\kt$,
redefine $\br x|{\cal Q}_j|y\kt$ to be the inverse Fourier
transform of $\br x|{\cal Q}_j|p\kt+u_+(p)\,e^{ipx}+
u_-(p)\,e^{-ipx}$ over $p$, fix $u_\pm$ in such a way that $\br
x|{\cal Q}_j|y\kt^*=\br y|{\cal Q}_j|x\kt$, and relabel $\br
x|{\cal Q}_j|p\kt+u_+(p)\,e^{ipx}+ u_-(p)\,e^{-ipx}$ by $\br
x|{\cal Q}_j|p\kt$.\footnote{One can show that there is always a
pair of functions $u_\pm$ that satisfy $\br x|{\cal
Q}_j|y\kt^*=\br y|{\cal Q}_j|x\kt$.}
    \item[({\em iv})] Let ${\cal Q}_j(x,p):=
\sqrt{2\pi}\,e^{-ixp}\br x|{\cal Q}_j|p\kt$ and order the terms in
${\cal Q}_j(x,p)$ in such a way that all $x$'s are placed to the
left of $p$'s. This yields ${\cal Q}_j$, if $x$ and $p$ are
identified with the corresponding operators.
    \end{itemize}
This prescription produces an expression for the operator ${\cal
Q}_j$ that is not manifestly Hermitian. To obtain a manifestly
Hermitian expression, one must use the standard commutation
relations to make the necessary reordering of the terms in ${\cal
Q}_j$. A manifestly Hermitian expression that significantly
simplifies the comparison of different operators (or different
expressions for the same operator) is the one having the following
symmetric form: $g_0(x)+h_0(p)+\sum_{k=1}^\infty
s_k\,\{g_k(x),h_k(p)\}$, where for all $k\in\{0,1,2,\cdots\}$,
$g_k$ and $h_k$ are real-valued nonconstant functions, $s_k\in\R$,
and $\{\cdot,\cdot\}$ is the anticommutator, \cite{jpa05b,p66}.

\section{Metric Operators for $v=i\epsilon x^3$}

The imaginary cubic potential, $v(x)=i\epsilon x^3$, has a real,
positive, discrete spectrum \cite{bender-prl,dorey,shin}.
Therefore, according to the general results of \cite{p2p3}, it
admits a nonempty set of positive-definite metric operators
$\eta_+$. Because this potential is purely imaginary, it is an
ideal toy model to apply the general method developed in the
preceding section.

The calculation of $Q_j$ for the potential $v=i\epsilon x^3$ has
been considered in \cite{bbj-prd} where the authors obtain the
operators $Q_j$ that yield the metric operator $\eta_+$ associated
with the ${\cal CPT}$-inner product. They achieve this in two
steps: (1) They use the restriction imposed by the choice of the
${\cal CPT}$-inner product to infer that $Q_j$ must be even in $x$
and odd in $p$ and to obtain a more restricted set of operator
equations for $Q_j$ (This amounts to setting $Q_j=0$ for even
$j$.) (2) They adopt an appropriate ansatz for the form of $Q_j$
and determine its free coefficients by enforcing the operator
equations for $Q_j$.

In this section, we will use our analytic method to obtain the
most general $Q_j$ for the potential $v=i\epsilon x^3$. Our method
is more systematic, technically more convenient, and more general,
for it does not involve making any kind of ({\em a priori})
assumptions about the general structure or symmetries of $Q_j$. In
particular, it does not rely on making an ansatz for $Q_j$.
Indeed, we will see that the ansatz used in \cite{bbj-prd} misses
a large class of ${\cal CPT}$-inner products even in the first
order of perturbation theory.

As we have argued in the preceding section for purely imaginary
potentials such as $v=i\epsilon x^3$, the operators $Q_j$ have the
general form (\ref{Qj=}). In the following, we will first obtain
the structure of the functions $F_j$ and $K_j$ appearing in
(\ref{Qj=}) and then present our calculation of ${\cal Q}_j$ and
$Q_j$ for $j=1,2,3$.

\subsection{Calculation of $F_j$ and $K_j$}

By definition \cite{critique,cjp-04c,jpa04b,jpa05b}, the physical
observables $O$ are linear operators that are Hermitian with
respect to the inner product $\br\cdot,\cdot\kt_{\eta_+}$, i.e.,
they are $\eta_+$-pseudo-Hermitian operators \cite{p1}. We can
construct them in terms of linear operators $o$ that are Hermitian
with respect to the reference inner product $\br\cdot|\cdot\kt$
according to $O=\eta_+^{-1/2}o\,\eta_+^{1/2}$. For example, for
$o=x$ and $o=p$, we find the $\eta_+$-pseudo-Hermitian position
($X$) and momentum $(P)$ operators,
    \bea
    X&:=&\eta_+^{-1/2}x\,\eta_+^{1/2}=e^{Q/2}x\,e^{-Q/2}=
    x+\sum_{k=1}^\infty\frac{(-1)^k}{2^k k!}\,[x,Q]_k,
    \label{X=}\\
    P&:=&\eta_+^{-1/2}p\,\eta_+^{1/2}=e^{Q/2}p\,e^{-Q/2}=
    p+\sum_{k=1}^\infty\frac{(-1)^k}{2^k k!}\,[p,Q]_k,
    \label{P=}
    \eea
respectively, \cite{critique,jpa04b}. These equations show how the
operators $Q_j$ and in particular $F_j$ and $K_j$ enter the
expression for the basic physical observables of the theory.

Next, let $\ell\in\R^+$ and scale $x$ according to $x\to x/\ell$.
The requirements that the canonical commutation relation is not
affected by this scaling and that the Hamiltonian
$H=\frac{1}{2}p^2+i\epsilon x^3$ changes by a total scaling (so
that the scaling $x\to x/\ell$ has the non-physical effect of
changing units, as it should), we find that $p\to\ell p$ and
$\epsilon\to\ell^5\epsilon$.

In light of (\ref{expand}), it is not difficult to see that the
right-hand side of (\ref{X=}) involves terms of the form
$\epsilon^j[x,F_j(p)]=i\epsilon^jF_j'(p)$ and
$\epsilon^j[x,K_j(p)]{\cal P}=i\epsilon^jK_j'(p){\cal P}$, where a
prime means the first derivative. Therefore, if we demand $X$ to
have the same scaling properties as $x$, i.e., $X\to X/\ell$, we
find that $x\to x/\ell$ implies
    \be
    F_j(p)\to F_j(\ell p)=\ell^{-5j} F_j(p),~~~~~~~~
    K_j(p)\to K_j(\ell p)=\ell^{-5j} K_j(p).
    \label{scale1}
    \ee
In particular, we have $F_j(p)=p^{-5j}\lambda_j(p)$ and
$K_j(p)=p^{-5j}\theta_j(p)$ where $\lambda_j(\ell p)=
\lambda_j(p)$ and $\theta_j(\ell p)=\theta_j(p)$ for all
$\ell\in\R^+$. Furthermore, the assumption that {\em the quantum
theory under study does not involve a preassigned hidden length
scale} implies that $\lambda_j$ and $\theta_j$ must be
constants.\footnote{This argument is effectively equivalent to and
provides a conceptual interpretation for what the authors' of
\cite{bbj-prd} refer to as ``dimensional consistency''.} This
together with the requirement that $F_j$ be real-valued and $K_j$
be ${\cal PT}$-invariant then imply that $\lambda_j\in\R$ and
$\theta_j=i^{j}\kappa_j$ for some $\kappa_j\in\R$. Therefore, for
the imaginary cubic potential, we have for all $j\in\Z^+$:
    \be
    F_j(p)=\frac{\lambda_j}{p^{5j}},~~~~~~~
    K_j(p)=\frac{i^{j}\kappa_j}{p^{5j}},
    \label{restrict-FKj}
    \ee
where $\lambda_j,\kappa_j\in\R$ are arbitrary.\footnote{We can
also apply the assumption of lack of a hidden length scale in the
underlying classical theory to arrive at (\ref{restrict-FKj}).}

\subsection{Calculation of $Q_j$}

For the potential $v=i\epsilon x^3$, we have
$H_0=\frac{1}{2}\,p^2$, $H_1=ix^3$, and $R_1=-2ix^3$. Hence, the
most general form of $Q_1$ is obtained by setting $j=1$ in
(\ref{Qj=}), where $\br x|{\cal Q}_1|y\kt$ is a particular
solution of the wave equation
    \be
    [-\partial_x^2+\partial_y^2]\br x|Q_1|y\kt=
    -4ix^3\delta(x-y),
    \label{weq-1}
    \ee
and $\delta(x)$ denotes the Dirac delta function. To obtain ${\cal
Q}_1$ we use the prescription described in Section~2. Taking the
Fourier transform of (\ref{weq-1}) and solving (\ref{ode}) for a
particular solution, we find
    \be
    \br x|{\cal Q}_1|p\kt=\frac{1}{2\sqrt{2\pi}}\, e^{i p x}
    \left(x^4\frac{1}{p}+2i\,x^3\frac{1}{p^2}-3\,x^2\frac{1}{p^3}-
    3i\,x\,\frac{1}{p^4}\right).
    \label{Q1xp}
    \ee
Hence,
    \be
    {\cal Q}_1=\frac{1}{2}\left(x^4\frac{1}{p}+2i\,x^3\frac{1}{p^2}-
    3\,x^2\frac{1}{p^3}-3i\,x\,\frac{1}{p^4}\right).
    \label{Q1=1}
    \ee
We can evaluate the inverse Fourier transform of (\ref{Q1xp}) to
obtain
    \be
    \br x|{\cal Q}_1|y\kt=\frac{i}{8}\,x y(x^2+y^2)\,{\rm
    sign}(x-y),
    \label{Q1=}
    \ee
where ${\rm sign}(x):=x/|x|$ for $x\neq 0$ and ${\rm sign}(0):=0$.

Eq.~(\ref{Q1=}) shows that indeed ${\cal Q}_1$ is a Hermitian
operator. With the help of the identities
    \be
    [x,f(p)]=if'(p),~~~~~~~~~~[x^3,f(p)]=
    \frac{3i}{2}\,
    \{x^2,f'(p)\}+\frac{i}{2}\,f^{(3)}(p),
    \label{id1}
    \ee
where $f^{(n)}$ stands for the $n$-th derivative of $f$ and
$f':=f^{(1)}$, we can express ${\cal Q}_1$ in the following
manifestly Hermitian form
    \be
    {\cal Q}_1=\frac{1}{4}\,\{x^4,\frac{1}{p}\}+
    \frac{3}{4}\,\{x^2,\frac{1}{p^3}\}+\frac{3}{p^5}.
    \label{Q1=p}
    \ee

A useful check on the validity of our calculation of ${\cal Q}_1$
is to verify that (\ref{Q1=}) satisfies (\ref{weq-1}). We have
checked this by direct substitution of (\ref{Q1=}) in the
left-hand side of (\ref{weq-1}).

Substituting (\ref{Q1=p}) in (\ref{Qj=}) with $j=1$ and using
(\ref{restrict-FKj}) we find the most general form of $Q_1$,
namely
    \be
    Q_1=\frac{1}{4}\,\{x^4,\frac{1}{p}\}+
    \frac{3}{4}\,\{x^2,\frac{1}{p^3}\}+\frac{\tilde\lambda_1}{p^5}+
   \frac{i\kappa_1}{p^5}\,{\cal P},
    \label{gen-Q1-rest}
    \ee
where $\tilde\lambda_1:=\lambda_1+3$ and $\kappa_1$ are arbitrary
real constants.

Now, we are in a position to compare our result for $Q_1$ with
that obtained by Bender, Brody, and Jones (BBJ) \cite{bbj-prd},
namely
    \be
    Q_1^{\rm BBJ}=\frac{1}{32}\left(x^4\frac{1}{p}+4x^3\frac{1}{p}
    \,x+6x^2\frac{1}{p}\,x^2+4x\,\frac{1}{p}\,x^3+\frac{1}{p}\,x^4
    \right)+\frac{\alpha}{p},
    \label{Q1-cpt-1}
    \ee
where $\alpha\in\R$ is arbitrary. We can use the identities,
    \be
    x^3\,\frac{1}{p}\,x+x\,\frac{1}{p}\,x^3=\{x^4,\frac{1}{p}\}+
    3\{x^2,\frac{1}{p^2}\}+\frac{12}{p^5},~~~~~~~~
    x^2\,\frac{1}{p}\,x^2=\frac{1}{2}\,\{x^4,\frac{1}{p}\}+
    2\{x^2,\frac{1}{p^3}\}+\frac{12}{p^5},
    \label{2}
    \ee
to express (\ref{Q1-cpt-1}) in the form
    \be
    Q_1^{\rm BBJ}=\frac{1}{4}\,\{x^4,\frac{1}{p}\}+
    \frac{3}{4}\,\{x^2,\frac{1}{p^3}\}+(\alpha+\frac{15}{4})\,
    \frac{1}{p^5}.
    \label{Q1-CPT}
    \ee
Clearly, $Q_1^{\rm BBJ}$ defines a one-parameter subfamily of the
operators of the form (\ref{gen-Q1-rest}). It corresponds to the
choice $\tilde\lambda_1=\alpha+\frac{15}{4}$ and $\kappa_1=0$.

It is interesting to see that (\ref{gen-Q1-rest}) does also define
a ${\cal CPT}$-inner product, for it satisfies all the conditions
stated in \cite{bbj-prd}. In particular, if we use the number
operator $N=\frac{1}{2}(x^2+p^2-1)$ to express ${\cal P}$ in the
form\footnote{Recall that the eigenfunctions $\br x|n\kt$ of the
harmonic oscillator Hamiltonian $\frac{1}{2}(x^2+p^2-1)$ are even
(respectively odd) functions of $x$ for even (respectively odd)
values of the spectral label $n$. Hence ${\cal P}=(-1)^N$.}
    \be
    {\cal P}=(-1)^N=e^{i\pi N}=e^{\frac{i\pi}{2}(
    p^2+x^2-1)}=-ie^{\frac{i\pi}{2}\,(p^2+x^2)},
    \label{N=}
    \ee
we see that the last term in (\ref{gen-Q1-rest}) is also even in
$x$ and odd in $p$. The analysis of \cite{bbj-prd} seems to have
missed this term, because it relies on the choice of a particular
ansatz for $Q_1$.

Our derivation of $Q_1$ shows that, under the assumption that the
quantum system defined by the Hamiltonian
$H=\frac{p^2}{2}+i\epsilon\,x^3$ does not have a hidden length
scale, up to (and including) the first order terms in the
perturbation parameter $\epsilon$, the ${\cal CPT}$-inner products
are the most general inner products that restore the unitarity of
the dynamics. This result does not, however, extend to higher
orders in perturbation theory. It already fails in the second
order, because in general
    \be
    Q_2=F_2(p)+
    K_2(p){\cal P}=\frac{1}{p^{10}}\,
    (\lambda_2-\kappa_2\,{\cal P}),
    \label{Q2=explicit}
    \ee
whereas for a ${\cal CPT}$-inner product $Q_2=0$. This is a
manifestation of the fact that the ${\cal CPT}$-inner products
form a proper subset of the set of all allowed inner products.

Next, we wish to compute $Q_3$ using our method. We postpone the
details of this calculation to Appendix~B. Here we outline its
general strategy.

The first step in the calculation of $Q_3$ is the determination of
$R_3$ which in view of (\ref{e3}) is given by
    \be
    R_3=-\frac{i}{6}[[x^3,Q_1],Q_1].
    \label{R3=}
    \ee
We recall that according to the prescription explained in
Section~2 we need to compute $\br x|R_3|p\kt$. Therefore, instead
of evaluating the double commutator in (\ref{R3=}), we compute
    \be
    \br x|R_3|y\kt=-\frac{1}{6}\,\br x|[[ix^3,{Q}_1],
    {Q}_1]|y\kt=-\frac{i}{6}\int_{-\infty}^\infty dz
    (x^3+y^3-2z^3)\br x|{Q}_1|z\kt \br z|{Q}_1|y\kt,
    \label{R3xy}
    \ee
and take its Fourier transform over $y$.\footnote{The advantage of
this approach is that the above calculation can be done using
Mathematica.} Next, we insert $Q_1={\cal Q}_1+\lambda
p^{-5}+i\kappa p^{-5}{\cal P}$, which is equivalent to
(\ref{gen-Q1-rest}), in Eq.~(\ref{R3xy}) and use (\ref{Q1=}) to
perform the integral over $z$. Doing the necessary calculations,
as outlined in Appendix~B, we then find
    \be
    R_3=S_{0,0}+\lambda_1S_{1,0}+\kappa_1S_{0,1}+
    \lambda_1\kappa_1 S_{1,1}+\lambda_1^2 S_{2,0}+
    \kappa_1^2 S_{0,2},
    \label{R3xy=expand}
    \ee
where the operators $S_{\mu,\nu}$ are defined by Eqs.~(\ref{s00})
-- (\ref{s02}) below.

In view of (\ref{R3xy=expand}), we can construct a particular
solution of (\ref{e}) of the form
    \be
    {\cal Q}_3=T_{0,0}+\lambda_1T_{1,0}+\kappa_1T_{0,1}+
    \lambda_1\kappa_1 T_{1,1}+\lambda_1^2 T_{2,0}+
    \kappa_1^2 T_{0,2},
    \label{Q3xy=expand}
    \ee
where $T_{\mu,\nu}$ solves the operator equation
    \be
    [p^2,T_{\mu,\nu}]=2S_{\mu,\nu}.
    \label{op-weq-T}
    \ee
Hence, $\br x|T_{\mu,\nu}|y\kt$ is a particular solution of the
wave equation
    \be
    (-\partial_x^2+\partial_y^2)\br x|T_{\mu,\nu}|y\kt=
    2\br x|S_{\mu,\nu}|y\kt.
    \label{weq-T}
    \ee
In Appendix~B we construct such solutions and check that the
corresponding operators, $T_{\mu,\nu}$, are Hermitian. This shows
that the operator ${\cal Q}_3$, as given by (\ref{Q3xy=expand}),
is Hermitian. Consequently Eqs.~(\ref{restrict-FKj}) hold for
$j=3$.

Using (\ref{Qj=}), (\ref{restrict-FKj}), (\ref{Q3xy=expand}), and
the explicit form of the operators $T_{\mu,\nu}$ given in
Appendix~B, we obtain after a lengthy calculation
    \be
    Q_3=\sum_{\ell=1}^5 d_{0\ell}\:
    \{x^{2\ell},\frac{1}{p^{15-2\ell}}\}
    +\frac{\tilde\lambda_3}{p^{15}}-i\left(
    \sum_{\ell=1}^4
    d_{1\ell}\:\{x^{2\ell},\frac{1}{p^{15-2\ell}}\}+
    \frac{\tilde\kappa_3}{p^{15}}\right){\cal P},
    \label{Q3-final}
    \ee
where
    \bea
    d_{01}&:=&c_{001}+\lambda_1\,c_{101}+\lambda_1^2c_{201}+
    \kappa_1^2c_{021},\quad\quad\quad
    d_{02}:=c_{002}+\lambda_1\,c_{102}+\kappa_1c_{012}\nn\\
    d_{03}&:=&c_{003}+\lambda_1\,c_{103},\quad\quad\quad
    d_{04}=c_{004},\quad\quad\quad d_{05}=c_{005}\quad\quad\quad
    d_{11}:=\kappa_1(c_{011}+\lambda_1c_{111}),\nn\\
    d_{12}&:=&\kappa_1(c_{012}+\lambda_1c_{112}),\quad\quad\quad
    d_{13}:=\kappa_1c_{013},\quad\quad\quad
    d_{14}:=\kappa_1c_{014},\nn\\
    \tilde\lambda_3&:=&\lambda_3+2(c_{000}+\lambda_1c_{100}+
    \lambda_1^2c_{200}+\kappa_1^2c_{020}),\quad\quad\quad
    \tilde\kappa_3:=\kappa_3+2\kappa_1(c_{010}+\lambda_1c_{110}),
    \nn
    \eea
$c_{\mu\nu\ell}$ are real constants given in Table~\ref{tab3},
$\lambda_1,\kappa_1$ are the free real parameters determining
$Q_1$, and $\lambda_3,\kappa_3$ (or equivalently
$\tilde\lambda_3,\tilde\kappa_3$) are another pair of free real
parameters.

Note that the freedom in the choice of parameters $\alpha$ and
$\beta$ entering the expression obtained in \cite{bbj-prd} for
$Q_3$ corresponds to the freedom in the choice of $\lambda_1$ and
$\lambda_3$ (or $\tilde\lambda_3$). Furthermore, it is not
difficult to see that $Q_3$ constructed above fulfils all the
conditions imposed by the choice of ${\cal CPT}$-inner product.
The fact that the approach of \cite{bbj-prd} do not reveal the
presence of the terms proportional to $\kappa_1$ and
$\tilde\kappa_3$ stems from the particular choice for the ansatz
used in \cite{bbj-prd}.

We conclude this section by mentioning that we can similarly apply
our method to obtain the general form of $Q_j$ for $j>3$. But as
expected the algebra becomes quite involved.

\section{Physical Implications of Changing the Metric Operator
in Quantum and Classical Treatments}

In order to understand how the freedom in the choice of the metric
operator affects the physical content of the theory, we will use
the most general metric operator to perform a perturbative
calculation of the pseudo-Hermitian position (\ref{X=}) and
momentum (\ref{P=}) operators, the equivalent Hermitian
Hamiltonian,
    \be
    h:=\eta_+^{1/2}H\,\eta_+^{-1/2}=e^{-Q/2}H\,e^{Q/2},
    \label{herm-h}
    \ee
and the classical Hamiltonian\footnote{$x_c$ and $p_c$ appearing
in (\ref{Hc=}) are respectively the classical position and
momentum observables.},
    \be
    H_c(x_c,p_c):=\left.\lim_{\hbar\to 0}h(x,p)
    \right|_{x\to x_c,p\to p_c},
    \label{Hc=}
    \ee
for the system defined by the imaginary cubic potential.

We can easily show, using (\ref{bch}), (\ref{X=}), (\ref{P=}) and
(\ref{herm-h}), that
    \bea
    X&=&x-\frac{1}{2}\,[x,Q_1]\,\epsilon+
    {\cal O}(\epsilon^2),\quad\quad\quad\quad
    P=p-\frac{1}{2}\,[p,Q_1]\,\epsilon+
    {\cal O}(\epsilon^2),
    \label{XP-1st}\\
    h&=&\frac{p^2}{2}+\frac{i}{4}\,[x^3,Q_1]\,\epsilon^2+
    +\frac{i}{4}\,[x^3,Q_2]\,\epsilon^3+
    {\cal O}(\epsilon^4),
    \label{h=3rd}
    \eea
where ${\cal O}(\epsilon^n)$ stand for terms of order $n$ and
higher in powers of $\epsilon$. Therefore, we first use
(\ref{gen-Q1-rest}) and (\ref{Q2=explicit}) to compute
    \bea
    \left[x\,,\,Q_1\right]&=&-\frac{i}{4}\,\left(
    \{x^4,\frac{1}{p^2}\}+9\,\{x^2,\frac{1}{p^4}\}
    +\frac{20\tilde\lambda_1}{p^6}-
    4\kappa_1\,\{x,\frac{1}{p^5}\}\:{\cal P}\right),
    \label{e11}\\
    \left[p\,,\,Q_1\right]&=&-\frac{i}{2}\left(2\{x^3,\frac{1}{p}\}+
    3\{x,\frac{1}{p^3}\}-\frac{4\kappa_1}{p^4}\,{\cal P}\right),
    \label{e12}\\
    \left[x^3,Q_1\right]&=&-\frac{3i}{4}\left(\{x^6,\frac{1}{p^2}\}
    +22\{x^4,\frac{1}{p^4}\}+(510+10\tilde\lambda_1)
    \{x^2,\frac{1}{p^6}\}+\right.\nn\\
    &&\left.
    \hspace{2cm}
    \frac{8820+140\tilde\lambda_1}{p^8}
    -\frac{4}{3}\,\kappa_1\{x^3,\frac{1}{p^5}\}\:{\cal P}\right),
    \label{e13}\\
    \left[x^3,Q_2\right]&=&-15i\lambda_2\left(\{x^2,\frac{1}{p^{11}}\}
    +\frac{44}{p^{13}}\right)-\kappa_2\{x^3,\frac{1}{p^{10}}\}\:
    {\cal P}.
    \label{e14}
    \eea
Substituting these relations in (\ref{XP-1st}) and (\ref{h=3rd}),
we have
    \bea
    X&=&x+\frac{i}{8}\,\left(
    \{x^4,\frac{1}{p^2}\}+9\,\{x^2,\frac{1}{p^4}\}
    +\frac{20\tilde\lambda_1}{p^6}-
    4\kappa_1\,\{x,\frac{1}{p^5}\}\:{\cal P}\right)\epsilon+
    {\cal O}(\epsilon^2),
    \label{X=man}\\
    P&=&p+\frac{i}{4}\left(2\{x^3,\frac{1}{p}\}+
    3\{x,\frac{1}{p^3}\}-\frac{4\kappa_1}{p^4}\,{\cal P}\right)
    \epsilon+ {\cal O}(\epsilon^2),
    \label{P=man}\\
    h&=&\frac{p^2}{2}+\frac{3}{16}
    \left(\{x^6,\frac{1}{p^2}\}
    +22\{x^4,\frac{1}{p^4}\}+(510+10\tilde\lambda_1)
    \{x^2,\frac{1}{p^6}\}+\right.\nn\\
    &&\left.
    \hspace{.7cm}
    \frac{8820+140\tilde\lambda_1}{p^8}
    -\frac{4}{3}\,\kappa_1\{x^3,\frac{1}{p^5}\}\:{\cal P}\right)
    \epsilon^2+\nn\\
    &&\hspace{1.4cm}\frac{1}{4}\left(
    15\lambda_2 (\{x^2,\frac{1}{p^{11}}\}
    +\frac{44}{p^{13}} )-i\kappa_2\{x^3,\frac{1}{p^{10}}\}\:
    {\cal P}\right)\epsilon^3+
    {\cal O}(\epsilon^4).
    \label{h=man}
    \eea
Equations (\ref{X=man}) and (\ref{P=man}) show how the free
parameters $\tilde\lambda_1$ and $\kappa_1$, that determine the
metric operator up to terms of order $\epsilon$, enter the
definition of the basic observables of the theory.

The quantum theory defined by the imaginary cubic potential can be
described by the manifestly Hermitian Hamiltonian $h$ within the
framework of the standard quantum mechanics,
\cite{jpa-2003,cjp-04c}. However, as seen from Eq.~(\ref{h=man})
this equivalent Hermitian description is sensitive to the choice
of the metric. {\em The metric-independence of the Hamiltonian $H$
and the metric-dependence of the physical observables $O$ in the
non-Hermitian description of the quantum system are traded with
the metric-dependence of the Hamiltonian $h$ and the
metric-independence of the physical observables $o$ in its
Hermitian description.}

Having calculated the Hermitian Hamiltonian $h$, we can determine
the classical Hamiltonian (\ref{Hc=}). This requires making the
$\hbar$-dependence of the terms in (\ref{h=man}) explicit. We do
this by letting $h\to \ell^{-2} m^{-1}\hbar^2 h$,
$x\to\ell^{-1}x$, $p\to\ell\hbar^{-1}p$, where $\ell$ is an
arbitrary length scale and $m$ is the mass, \cite{jpa05b}. Making
this transformations, replacing the quantum observables with their
classical counterparts, i.e., $x\to x_c$ and $p\to p_c$, and
taking the limit $\hbar\to 0$, we find the following remarkably
simple expression for the classical Hamiltonian (\ref{Hc=}).
    \be
    H_c=\frac{p^2_c}{2m}+\frac{3}{8}\,m\epsilon^2\,\frac{x_c^6}{p_c^2}
    +{\cal O}(\epsilon^4).
    \label{H-classical}
    \ee
Clearly $H_c$ is an even and nonnegative function of $x_c$ and
$p_c$. This is an indication that it supports closed classical
phase space orbits. As shown in Figure~1, this is actually the
case.

A more important observation is that the terms in (\ref{h=man})
that involve $\tilde\lambda_1$, $\kappa_1$, $\lambda_2$ and
$\kappa_2$ do not contributes to the classical Hamiltonian,
because they involve positive integer powers of $\hbar$ that
vanish in the classical limit $\hbar\to 0$. It is not difficult to
see that the same behavior holds in all orders of perturbation;
the terms involving $\lambda_j$ and $\kappa_j$, that characterize
the freedom in the choice of the metric operator, do not
contribute to the classical Hamiltonian. This means, at least for
the system we consider, that {\em the classical Hamiltonian is not
sensitive to the choice of the metric operator.} We expect this
assertion to hold true generally.

\begin{figure}
\centerline{\epsffile{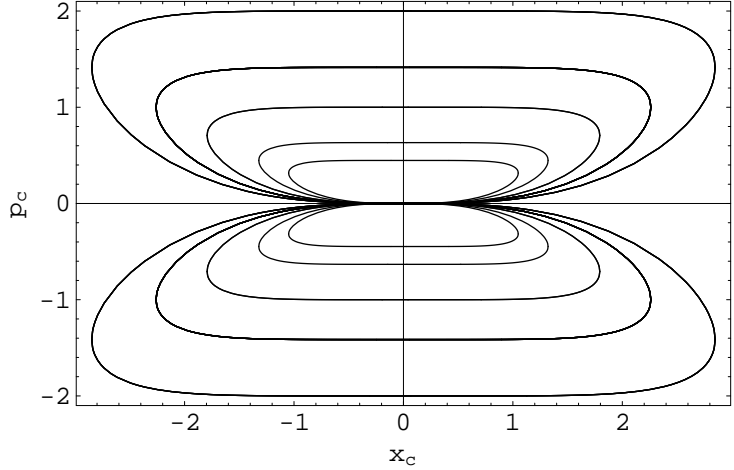}} \caption{Typical phase-space
orbits for the classical Hamiltonian $H_c$, for $\epsilon=0.1$ and
$m=1$.} \label{fig1}
\end{figure}

\section{Pseudo-Hermitian Quantization of the Free Particle}

If we set $H_0=p^2/2$ and $v_0=H_1=0$ in the constructions of
Section~2, we find the most general metric operator
$\eta_+=e^{-Q}$ that renders the free particle Hamiltonian $H=H_0$
Hermitian. It is not difficult to see that in this case $R_j=0$
for all $j\in\Z^+$ and that
    \be
    Q=F +K {\cal P},
    \label{Q-free}
    \ee
for a pair of functions $F$ and $K$ of $p$. Again the fact that
the system does not involve a particular length scale implies that
$F(p)=:\lambda$ and $K(p)=:\kappa$ are real constants. Hence
    \be
    \eta_+=e^{-\lambda}e^{-\kappa {\cal P}}=
    e^{-\lambda}[\cosh(\kappa)-\sinh(\kappa){\cal P}].
    \label{free-eta}
    \ee

For this system the pseudo-Hermitian position (\ref{X=}) and
momentum (\ref{P=}) operators take the form
    \be
    X=x\:e^{-\kappa {\cal P}}=e^{\kappa {\cal P}}x,~~~~~~
    P=p\:e^{-\kappa {\cal P}}=e^{\kappa {\cal P}}p.
    \label{XP-free}
    \ee
Clearly the Hermitian hamiltonian $h$ and the classical
Hamiltonian $H_c$ coincide with $H$ and $p_c^2/2$, respectively.
The pseudo-Hermitian quantization scheme defined by
    \be
    x_c\to X~~~~~~p_c\to P,~~~~~~~~
    \mbox{Poission bracket~$\to~-i\,$commutator},
    \label{ph-quantize}
    \ee
is the most general physically admissible canonical quantization
of the free particle in one dimension that is compatible with the
usual form of the free particle Hamiltonian. Note that for
$\kappa\neq 0$ the operators $X$ and $P$ are not Hermitian with
respect to the usual $L^2$-inner product. Yet they furnish a
unitary irreducible representation of the Heisenberg-Weyl algebra
$[x,p]=i$. The representation space is the Hilbert space ${\cal
H}_{\rm phys}=\{\psi:\R\to\C| \br \psi,\psi\kt_{\eta_+}<\infty\}$
endowed with the inner product,
    \be
    \br\phi,\psi\kt_{\eta_+}:=\br\phi|\eta_+|\psi\kt=
    e^{-\lambda/2}
    \cosh(\kappa)\int_{-\infty}^\infty dx\:\phi(x)^*\psi(x)-
    e^{-\lambda/2}
    \sinh(\kappa)\int_{-\infty}^\infty dx\:\phi(x)^*\psi(-x),
    \label{ph-inn}
    \ee
where $\psi,\phi:\R\to\C$ are arbitrary
functions.\footnote{Because the metric operator (\ref{free-eta})
is a bounded operator acting in $L^2(\R)$, ${\cal H}_{\rm phys}$
and $L^2(\R)$ are identical as vector spaces. What distinguishes
them from one another is their inner products.}

As suggested by (\ref{XP-free}) and (\ref{ph-inn}), the
pseudo-Hermitian quantization of the free particle may be viewed
as a smooth deformation of the standard (Hermitian) canonical
quantization (which corresponds to taking $\kappa=0$).

Next, we recall that $\eta_+^{1/2}$ defines a unitary operator
mapping ${\cal H}_{\rm phys}$ onto $L^2(\R)$,
\cite{jpa-2003,jpa04b,jpa05b}. Under this mapping $X$ and $P$ are
transformed to $x$ and $p$ respectively and $H$ is left invariant.
Therefore, the pseudo-Hermitian quantum systems associated with
all $\lambda,\kappa\in\R$ are physically equivalent. In
particular, there is no physical reason to choose $\kappa=0$. We
should like to warn however that taking $\kappa\neq 0$ has unusual
yet completely consistent consequences. For example, the localized
states are described by wave functions that in general have two
delta function singularities! Up to an irrelevant phase factor,
the delta-function normalized localized state vector centered at
$y\in\R$ is given by $|\xi^{(y)}\kt:=\eta_+^{-1/2}|y\kt$,
\cite{jpa04b,jpa05b}. Therefore, according to (\ref{free-eta}),
    \[\br x|\xi^{(y)}\kt=e^{\lambda/2}[
    \cosh(\mbox{$\frac{\kappa}{2}$})\,\delta(x-y)+
    \sinh(\mbox{$\frac{\kappa}{2}$})\,\delta(x+y)].\]
Note however that this is not the physical position wave function
for the localized state. The latter has the form $\br\xi^{(x)},
\xi^{(y)}\kt_{\eta_+}=\br x|y\kt=\delta(x-y)$.

Finally, we would like to point out that according to
(\ref{XP-free}), $X^2=x^2$, $XP=xp$, $P^2=p^2$. Therefore if we
perform the pseudo-Hermitian canonical quantization
(\ref{ph-quantize}) on any classical Hamiltonian $H_c$ that can be
constructed out of $x_c^2$, $x_cp_c$ and $p_c^2$, we obtain the
same expression for the quantum Hamiltonian as the one obtained
using the usual canonical quantization. Conversely, every quantum
Hamiltonian $H$ that can be expressed in terms of $x^2$, $xp+px$,
and $p^2$ admits a metric operator of the form (\ref{free-eta}).

\section{Higher Dimensions and Scaling-Invariance}

The situation is considerably more complicated in $n$-dimensional
configuration spaces with $n>1$. In this case the operator
equations (\ref{e}) correspond to certain ultrahyperbolic
equations in $(n+n)$-dimensions. These equations have peculiar
properties, and except for very special cases their standard
Cauchy problem is not well-posed \cite{tegmark}. For the free
particle in $n$-dimensions, $Q$ satisfies
    \be
    (-\nabla_x^2+\nabla_y^2)\br\vec x|Q|\vec y\kt=0,
    \label{ultra}
    \ee
where $\vec x=(x_1,x_2,\cdots,x_n)$ and
$\nabla_x^2:=\sum_{i=1}^n\partial_{x_i}^2$. It is not difficult to
see that (\ref{Q-free}) does not provide the most general solution
of (\ref{ultra}).\footnote{For example, for $n=2$, the angular
momentum operator, $Q=x_1p_2-x_2p_1$, which is clearly not of the
form (\ref{Q-free}), provides a solution of (\ref{ultra}).} It
only provides a special class of solutions. An important
distinction between this class for $n>1$ and the corresponding
general solution for $n=1$ is that for $n>1$ the requirement that
the theory must not depend on an arbitrary length scale does not
restrict $F$ and $K$ to constants. This is simply because one can
now use the ratios of the components $p_i$ of the momentum
operator $\vec p$ to construct scaling-independent operators,
i.e., require that $F$ and $K$ are functions of the homogeneous
$p$-coordinates. For example for $n=2$, we can take $F=F(p_1/p_2)$
and $K=K(p_1/p_2)$.

A similar situation occurs in one-dimension provided that the
system involves an $\epsilon$-independent length scale. A simple
example is the scattering potential,
    \be
    v(x)=\left\{\begin{array}{ccc}
    -i\epsilon\:{\rm sign}(x)&{\rm for}& |x|<\frac{L}{2}\\
    &&\\
    0&{\rm for}& |x|\geq\frac{L}{2},
    \end{array}\right.
    \label{scatter}
    \ee
which includes the length scale $L$. This potential has a real
spectrum \cite{p66}. It is also imaginary. Therefore, we can apply
our method to obtain the general form of $Q_j$. In \cite{p66} we
have used a considerably more difficult spectral method to obtain
the following particular solution of (\ref{weq}).
    \be
    {\cal Q}_1(x,y)=\br x|{\cal Q}_1|y\kt=
    {\mbox{$\frac{iL}{16}$}}\,
    (|x+y+L|+|x+y-L|-2|x+y|-2L)\:{\rm sign}(x-y).
    \label{scatter-Q1}
    \ee
According to the results of Section~2, the general form of $Q_1$
is given by (\ref{Qj=}) with $j=1$. However, unlike the case of
imaginary cubic potential, the operators $F_j$ and $K_j$ appearing
in (\ref{Qj=}) can now be complicated functions of $Lp$.
Specifically, given the fact that for the system defined by the
potential (\ref{scatter}), the scaling transformation $x\to
x/\ell$ induces the transformation $\epsilon\to\ell^2\epsilon$, it
is not difficult to infer that $F_j$ and $K_j$ have the form
    \[F_j=\frac{\lambda_j(Lp)}{p^{2j}},~~~~~~~~~
    K_j=\frac{\kappa_j(Lp)}{p^{2j}},\]
where $\lambda_j$ and $\kappa_j$ are respectively real-valued and
${\cal PT}$-invariant functions. In particular, we have
    \bea
    Q_{1}&=&{\cal Q}_1+p^{-2}\,[\lambda_{1}(Lp)+
    \kappa_{1}(Lp)\:{\cal P}],
    \label{scatter-Q1k}
    \eea
where ${\cal Q}_1$ is given by (\ref{scatter-Q1}) and
    \bea
    Q_{2}&=&p^{-4}\,[\lambda_{2}(Lp)+
    \kappa_{2}(Lp)\:{\cal P}].
    \label{scatter-Q2k}
    \eea
The derivation of (\ref{scatter-Q2k}) using the spectral approach
pursued in \cite{p66} is an extremely difficult (and open)
problem.

\section{Concluding Remarks}

In this paper we offer a method to compute the most general metric
operator for a given quasi-Hermitian standard Hamiltonian in
one-dimension. This method can be conveniently used for imaginary
potentials. It allows a better understanding of the issue of
non-uniqueness of the metric operator and the positive-definite
inner products that render the quantum theory unitary.

Applying this method to the imaginary cubic potential we
discovered a new class of ${\cal CPT}$-inner products and
established the existence of admissible non-${\cal CPT}$-inner
products. Furthermore, we showed that the freedom in the choice of
the metric operator affected the basic physical observables of the
theory in its non-Hermitian representation and the Hamiltonian in
its Hermitian representation. An interesting consequence of our
investigation is that the underlying classical system is not
sensitive to the choice of the inner product.

We also used our method to obtain a complete characterization of
pseudo-Hermitian quantization of the free particle Hamiltonian in
one-dimension.

We wish to emphasize that the choice of the metric operator can be
related to that of an irreducible set of observables as explained
in \cite{geyer2}. Specifically, the requirement that the members
of an irreducible set of operators\footnote{A set of operators
acting in a Hilbert space ${\cal H}$ is called irreducible if
there is no proper subspace of ${\cal H}$ that is left invariant
under the action of all members of this set.} be Hermitian fixes
the metric operator (up to an irrelevant constant coefficient)
\cite{geyer2,ashtekar}. The application of this program for ${\cal
PT}$-symmetric Hamiltonians having a real spectrum involves
selecting sufficiently many operators (all having real spectra)
such that together with the Hamitonian they form an irreducible
set. These operators must however be compatible in the sense that
there must exist a positive-definite metric operator $\eta_+$ such
that all of them be $\eta_+$-pseudo-Hermitian. In general the
determination of the compatibility of the operators that are to be
chosen is a difficult task. In particular it requires the
knowledge of the most general positive-definite metric operator
that renders the Hamiltonian pseudo-Hermitian. Therefore, the
results we have reported in this paper play a central role in
employing the method of \cite{geyer2} in ${\cal PT}$-symmetric
quantum mechanics. \vspace{.5cm}

{\bf End Note}: For the imaginary cubic potential considered here
it is sometimes argued that $\br x|\eta|y\kt$ is not an analytic
function of $\epsilon$ and its series expansion studied in
\cite{bbj-prd} and here is meaningless, because $\epsilon$ is not
dimensionless and scaling $x$ and $y$ one can make the numerical
value of $\epsilon$ as large or small as one wishes. This argument
is inconclusive, for it certainly applies to $e^{\epsilon x^5}$,
which is an analytic function of $\epsilon$ for all $x\in\R$, and
to $(1+\epsilon x^5)^{-1}$ which is analytic in $\epsilon$ for
$|x|<|\epsilon|^{1/5}$. As these examples suggest, in general
there is a region in the $x$-$y$ plane in which $\br x|\eta|y\kt$
admits a power series expansion in $\epsilon$. The issue of the
summability of this series is an extremely difficult open problem.
The fact that $\epsilon$ is not dimensionless implies that an
approximate evaluation of $\eta$ that is based on a truncation of
the power series for $\br x|\eta|y\kt$ will be valid only within a
region ${\cal D}_\epsilon$ of the $x$-$y$ plane whose size depends
on $\epsilon$. Obviously, because the contribution of every order
in $\epsilon$ to $\eta$ is dimensionless, scaling $\epsilon$ would
scale the size of ${\cal D}_\epsilon$.

\np
\section*{Appendix A: Derivation of the Operator Equations
Satisfied by $Q_j$}

In order to derive~(\ref{e}) we first substitute (\ref{H=HH}) and
(\ref{log}) in (\ref{ph}) and use (\ref{bch}) and the fact that
$H_1^\dagger=-H_1$ to obtain
    \be
    \epsilon H_1=-\frac{1}{2}\,
    \left(\dl H_0,Q\dr_{_1}+\dl \epsilon H_1,Q\dr_{_1}\right)
    \label{app-a1}
    \ee
where
    \be
    \dl A,Q\dr_{_1}:=e^{-Q}A\,e^{Q}-A=\sum_{k=1}^\infty
    \frac{1}{k!}\,[A,Q]_{_k}.
    \label{app-a2}
    \ee
Next, we substitute the right-hand side of Eq.~(\ref{app-a1}) for
the $\epsilon H_1$ in the term $\dl \epsilon H_1,Q\dr_1$ that
appears in this equation and do the same in the right-hand side of
the resulting equation. Repeating this procedure $\ell-1$ times
yields
    \be
    \epsilon H_1=\frac{1}{2}\sum_{m=1}^\ell \sum_{k=m}^\infty
    q_{mk}[H_0,Q]_{_k}+\frac{(-1)^\ell}{2^\ell}\,\dl
    \epsilon H_1,Q\dr_{_\ell},
    \label{pert}
    \ee
where $\ell\in\Z^+$ is arbitrary and for all $k,m\in\Z^+$
    \be
    q_{mk}:=\sum_{n=1}^m\frac{(-1)^n n^k}{k! 2^{m-1}}
    \mbox{\small$\left(\!\!\begin{array}{c} m\\n\end{array}\!\!
    \right)$},~~~~~~\mbox{\small $\left(\!\!\begin{array}{c}
    m\\n\end{array}\!\!\right)$}:=\frac{m!}{n!(m-n)!},~~~~~~
    \dl A,Q\dr_{_{k+1}}:=\dl\,\dl A,Q\dr_{_k},Q\dr_{_1}.
    \label{app-a3}
    \ee

In view of (\ref{expand}), (\ref{app-a2}), (\ref{pert}), and
(\ref{app-a3}), it is easy to see that
    \be
    \epsilon H_1=\frac{1}{2}\sum_{m=1}^\ell \sum_{k=m}^\ell
    q_{mk}[H_0,Q]_{_k}+{\cal O}(\epsilon^{\ell+1}),
    \label{pert2}
    \ee
where ${\cal O}(\epsilon^{\ell})$ stands for terms of order $\ell$
or higher in powers of $\epsilon$. We can employ the identity
$\sum_{m=1}^\ell \sum_{k=m}^\ell=\sum_{k=1}^\ell \sum_{m=1}^k$ to
express (\ref{pert2}) in the form
    \be
    \epsilon H_1=\frac{1}{2}\sum_{k=1}^\ell q_k[H_0,Q]_{_k}+
    {\cal O}(\epsilon^{\ell+1}),
    \label{pert2-new}
    \ee
where $q_k:=\sum_{m=1}^kq_{mk}$.

Next, we substitute (\ref{expand}) in(\ref{pert2-new}) and require
that (\ref{pert2-new}) be satisfied at each order $j\leq\ell$ of
the perturbation. For $j=1$, this implies
    \be
    H_1=-\frac{1}{2}\,[H_0,Q_1],
    \label{app-b-R1}
    \ee
where we have used $q_1=-1$. For $j\geq 2$, we solve for
$\epsilon^j[H_0,Q_j]$ in terms of the remaining terms of order
$\epsilon^j$ in (\ref{pert2-new}). This together with
(\ref{app-b-R1}) yield (\ref{e}).

\section*{Appendix B: Calculation of $Q_3$}

In view of (\ref{R3xy}), the operator $S_{\mu,\nu}$, that
determine $R_3$ according to (\ref{R3xy=expand}), are given by
    \bea
    \br x|S_{0,0}|y\kt&:=&-\frac{i}{6}\int_{-\infty}^\infty dz
    (x^3+y^3-2z^3)\br x|{\cal Q}_1|z\kt \br z|{\cal Q}_1|y\kt,
    \label{s00}\\
    \br x|S_{1,0}|y\kt&:=&-\frac{i}{6}\int_{-\infty}^\infty dz
    (x^3+y^3-2z^3)[\br x|{\cal Q}_1|z\kt \br z|p^{-5}|y\kt+
    \br x|p^{-5}|z\kt \br z|{\cal Q}_1|y\kt],
    \label{s10}\\
    \br x|S_{0,1}|y\kt&:=&\frac{1}{6}\int_{-\infty}^\infty dz
    (x^3+y^3-2z^3)[\br x|{\cal Q}_1|z\kt \br z|p^{-5}|-y\kt+
    \br x|p^{-5}|-z\kt \br z|{\cal Q}_1|y\kt],
    \label{s01}\\
    \br x|S_{1,1}|y\kt&:=&\frac{1}{6}\int_{-\infty}^\infty dz
    (x^3+y^3-2z^3)[\br x|p^{-5}|z\kt \br z|p^{-5}|-y\kt+
    \br x|p^{-5}|-z\kt \br z|p^{-5}|y\kt],
    \label{s11}\\
    \br x|S_{2,0}|y\kt&:=&-\frac{i}{6}\int_{-\infty}^\infty dz
    (x^3+y^3-2z^3) \br x|p^{-5}|z\kt \br z|p^{-5}|y\kt,
    \label{s20}\\
    \br x|S_{0,2}|y\kt&:=&-\frac{1}{6}\int_{-\infty}^\infty dz
    (x^3+y^3-2z^3) \br x|p^{-5}|-z\kt \br z|p^{-5}|-y\kt
    \label{s02}
    \eea
We can use (\ref{Q1=}) and the identities
    \be
    {\rm sign}(x)=\frac{1}{i\pi}\,
    \int_{-\infty}^\infty dk\:\frac{e^{ixk}}{k},~~~~~~~
    \br x|\frac{1}{p^5}|y\kt=
    \frac{1}{2\pi}\int_{-\infty}^\infty dk\:
    \frac{e^{i(x-y)k}}{k^5}=\frac{i}{48}\,(x-y)^4\:
    {\rm sign}(x-y),
    \label{id-sgn}
    \ee
to evaluate the integrals appearing in (\ref{s00}) -- (\ref{s02})
by turning them into a pair of Fourier transforms that we can
easily perform using Mathematica. The result is
    {\small
    \bea
    \br x|S_{0,0}|y\kt&=&-\frac{i}{26880}\,\left[
    15(x^{11}y-x\,y^{11})+7(x^9y^3-x^3y^9)-48(x^8y^4-x^4y^8)\right]
    {\rm sign}(x-y),
    \label{s00=}\\
    \br x|S_{0,1}|y\kt&=&\frac{1}{319334400}\,(x+y)^5\,
    \left[3115(x^7-y^7)-3818(x^6y-xy^6)+\right.\nn\\
    &&\hspace{2.5cm}\left.3120(x^5y^2-x^2y^5)
    -12123(x^4y^3-x^3y^4)\right]\,{\rm sign}(x+y),
    \label{s01=}\\
    \br x|S_{1,0}|y\kt&=&-\frac{i}{6386688}\,(x-y)^7\,\left[
    623(x^5+y^5)+1970(x^4y+xy^4)+\right.\nn\\
    &&\hspace{5cm}\left.3743(x^3y^2+x^2y^3)\right]\,
    {\rm sign}(x-y),
    \label{s10=}\\
    \br x|S_{1,1}|y\kt&=&\frac{1}{47900160}\,
    (x+y)^9\,[7(x^3-y^3)-15(x^2y-xy^2)]\,{\rm sign}(x+y),
    \label{s11=}\\
    \br x|S_{0,2}|y\kt&=&\frac{i}{95800320}\,
    (x-y)^9\,[29(x^3+y^3)+15(x^2y+xy^2)]\,{\rm sign}(x-y),
    \label{s02=}\\
    \br x|S_{2,0}|y\kt&=&\frac{i}{6386688}\, (x-y)^{11}\,(x+y)\,
    {\rm sign}(x-y).
    \label{s20=}
    \eea}
These equations show that, as expected, $S_{\mu,\nu}$ are
anti-Hermitian.

Next, we use (\ref{s00=}) -- (\ref{s20=}) and the the prescription
given in Section~2 to transform the wave equations (\ref{weq-T})
into the ordinary differential equation $(\partial_x^2+p^2)\br
x|T_{\mu,\nu}|p\kt=-2\br x|S_{\mu,\nu}|p\kt$, obtain particular
solutions of this equation, and determine the corresponding
operators $T_{\mu,\nu}$ that solve (\ref{op-weq-T}). This yields
    {\small
    \bea
    T_{0,0}&=&-\sum_{\ell=1}^{10} (-i)^\ell\, a_{00\ell}\:
    x^\ell\,\frac{1}{p^{15-\ell}},~~~~~~~
    T_{0,1}=\sum_{\ell=1}^{8} (-i)^{\ell+1}\,
    a_{01\ell}\: x^\ell\,\frac{1}{p^{15-\ell}}~{\cal P},
    \label{T00=}\\
    T_{1,0}&=&-\sum_{\ell=1}^{6} (-i)^\ell\, a_{10\ell}\:
    x^\ell\,\frac{1}{p^{15-\ell}},~~~~~~~
    T_{1,1}=\sum_{\ell=1}^{4} (-i)^{\ell+1}\,
    a_{11\ell}\: x^\ell\,\frac{1}{p^{15-\ell}}~{\cal P},
    \label{T11=}\\
    T_{0,2}&=&\sum_{\ell=1}^{4} (-i)^\ell\, a_{02\ell}\:
    x^\ell\,\frac{1}{p^{15-\ell}},~~~~~~~~~
    T_{2,0}=\sum_{\ell=1}^{2} (-i)^\ell\, a_{20\ell}\:
    x^\ell\,\frac{1}{p^{15-\ell}},
    \label{T20=}
    \eea}
where $a_{\mu\nu\ell}$ are real and positive constants given in
Table~\ref{tab1}.

Next, we wish to determine if $T_{\mu,\nu}$ and consequently the
operator ${\cal Q}_3$ given by (\ref{Q3xy=expand}) are Hermitian.
We can easily answer this question by inspecting the matrix
elements:
    \bea
    \br x|T_{0,0}|y\kt&=&i(x-y)^4\sum_{\ell=1}^{5}
    b_{00\ell}\: (x^\ell y^{10-\ell}+x^{10-\ell}y^\ell)
    \;{\rm sign}(x-y),
    \label{t00=}\\
    \br x|T_{0,1}|p\kt&=&(x+y)^6\sum_{\ell=1}^{4}
    b_{01\ell}\: (x^\ell y^{8-\ell}+x^{8-\ell}y^\ell)
    \;{\rm sign}(x+y),
    \label{t01=}\\
    \br x|T_{1,0}|y\kt&=&i(x-y)^8\sum_{\ell=1}^{3}
    b_{10\ell}\: (x^\ell y^{6-\ell}+x^{6-\ell}y^\ell)
    \;{\rm sign}(x-y),
    \label{t10=}\\
    \br x|T_{1,1}|y\kt&=&(x+y)^{10}\sum_{\ell=1}^{2}
    b_{11\ell}\: (x^\ell y^{4-\ell}+x^{4-\ell}y^\ell)
    \;{\rm sign}(x+y),
    \label{t11=}\\
    \br x|T_{0,2}|y\kt&=&i(x-y)^{10}\sum_{\ell=1}^{2}
    b_{02\ell}\: (x^\ell y^{4-\ell}+x^{4-\ell}y^\ell)
    \;{\rm sign}(x-y),
    \label{t02=}\\
    \br x|T_{2,0}|y\kt&=&i\,b_{201}\:(x-y)^{12}xy\;{\rm
sign}(x-y),
    \label{t20=}
    \eea
where $b_{\mu\nu\ell}$ are the real coefficients listed in
Table~\ref{tab2}. Eqs.~(\ref{t00=}) -- (\ref{t20=}) show that
indeed $T_{\mu,\nu}$ and consequently ${\cal Q}_3$ are Hermitian
operators. Hence in view of (\ref{Qj=}), (\ref{restrict-FKj}) and
(\ref{Q3xy=expand}), $Q_3$ has the following general form
    \be
    Q_3=T_{0,0}+\lambda_1T_{1,0}+\kappa_1T_{0,1}+
    \lambda_1\kappa_1 T_{1,1}+\lambda_1^2 T_{2,0}+
    \kappa_1^2 T_{0,2}+\lambda_3\,p^{-15}-i\kappa_3p^{-15}\:{\cal P},
    \label{Q3=cubic}
    \ee
where $\lambda_1$ and $\kappa_1$ are the real free parameters that
fix $Q_1$ and $\lambda_3$ and $\kappa_3$ are a pair of arbitrary
real constant coefficients.

In order to derive a manifestly Hermitian expression for $Q_3$, we
first derive, after a lengthy calculation, the following
manifestly Hermitian form of $T_{\mu,\nu}$.
    \be
    T_{\mu,\nu}=\left\{\begin{array}{ccc}
    \sum_{\ell=0}^{5} c_{\mu\nu\ell}\,\{x^{2\ell},
    \frac{1}{p^{15-2\ell}}\}&~~~~{\rm for}~~~~&\nu\neq 1\\
    &&\\
    -i\sum_{\ell=0}^{4} c_{\mu 1\ell}\,\{x^{2\ell},
    \frac{1}{p^{15-2\ell}}\}\:{\cal P}&~~~~{\rm for}~~~~&\nu=1,
    \end{array}\right.
    \label{Tab=manifest}
    \ee
where $c_{\mu\nu\ell}$ are real constants given in
Table~\ref{tab3}. Substituting (\ref{Tab=manifest}) in
(\ref{Q3=cubic}) we finally find (\ref{Q3-final}).

\np
{
}

\begin{table}[p]
    \vspace{.5cm}
\begin{center}
\begin{tabular}{||c||c|c|c|c|c|c||}
\hline \hline
 $\ell$ & 0 & 1 & 2 & 3 & 4 & 5
\\
\hline\hline
& & & & & & \\
$c_{00\ell}$&
    $\frac{141274966833}{32}$&
    $\frac{3830434839}{64}$&
    $\frac{23858793}{64}$&
    $\frac{43479}{32}$&
    $\frac{267}{64}$&
    $\frac{1}{80}$\\
& & & & & &\\
\hline & & & & & & \\
$c_{01\ell}$&
    $\frac{24081603}{20}$&
    $\frac{328947}{40}$&
    $\frac{16327}{80}$&
    $\frac{35}{48}$&
    $\frac{1}{480}$&
    -\\
& & & & & &\\
\hline & & & & & & \\
$c_{10\ell}$&
    $\frac{54563145}{16}$&
    $\frac{1430535}{16}$&
    $\frac{8695}{16}$&
    $\frac{5}{3}$&
    \scriptsize{$0$}&
    \scriptsize{$0$}\\
& & & & & &\\
\hline & & & & & & \\
$c_{11\ell}$&
    $\frac{1547}{4}$&
    $\frac{61}{4}$&
    $\frac{1}{12}$&
    \scriptsize{$0$}&
    \scriptsize{$0$}&
    -\\
& & & & & &\\
\hline & & & & & & \\
$c_{02\ell}$&
    \scriptsize{$-357$}&
    \scriptsize{$9$}&
    $\frac{1}{12}$&
    \scriptsize{$0$}&
    \scriptsize{$0$}&
    \scriptsize{$0$}\\
& & & & & & \\
\hline
& & & & & &\\
$c_{20\ell}$&
    $\frac{-2275}{4}$&
    $\frac{-25}{4}$&
    \scriptsize{$0$}&
    \scriptsize{$0$}&
    \scriptsize{$0$}&
    \scriptsize{$0$}\\
& & & & & &\\
\hline \hline\end{tabular}
  \end{center}
    \centerline{\parbox{13cm}{\caption{Numerical values of the
    coefficients $c_{\mu\nu\ell}$}
    \label{tab3}}}
    \end{table}

\begin{table}[p]
    \vspace{.5cm}
\begin{center}
\begin{tabular}{||c||c|c|c|c|c|c|c|c|c|c||}
\hline \hline
$\ell$ & 1 & 2 & 3 & 4 & 5 & 6 & 7 & 8 & 9 & 10\\
\hline\hline &  &  &  &  &  &  &  &  &  &
\\
$a_{00\ell}$&
    $\frac{2745171}{32}$ &
    $\frac{2745171}{32}$ &
    $\frac{677457}{16}$ &
    $\frac{439857}{32}$ &
    $\frac{52029}{16}$ &
    $\frac{9375}{16}$ &
    $\frac{651}{8}$ &
    $\frac{273}{32}$ &
    $\frac{5}{8}$ &
    $\frac{1}{40}$\\&  &  &  &  &  &  &  &  &  &
\\
\hline&  &  &  &  &  &  &  &  &  &
\\
 $a_{01\ell}$&
    $\frac{70317}{5}$ &
    $\frac{23592}{5}$ &
    $\frac{20207}{20}$ &
    $\frac{794}{5}$ &
    $\frac{777}{40}$ &
    $\frac{217}{120}$ &
    $\frac{7}{60}$ &
    $\frac{1}{240}$ &
    - &
    - \\&  &  &  &  &  &  &  &  &  &
\\
\hline &  &  &  &  &  &  &  &  &  &
\\
$a_{10\ell}$&
    $\frac{1110915}{8}$ &
    $\frac{363315}{8}$ &
    $\frac{36355}{4}$ &
    $\frac{9305}{8}$ &
    {\scriptsize$90$} &
    $\frac{10}{3}$ &
    -&
    -&
    -&
    -\\
&  &  &  &  &  &  &  &  &  &
\\
\hline &  &  &  &  &  &  &  &  &  &
\\
$a_{11\ell}$&
    $\frac{351}{2}$ &
    $\frac{71}{2}$ &
    $\frac{11}{3}$ &
    $\frac{1}{6}$ &
    -&
    -&
    -&
    -&
    -&
    -\\
&  &  &  &  &  &  &  &  &  &
\\
\hline &  &  &  &  &  &  &  &  &  &
\\
$a_{02\ell}$&
    \scriptsize{$388$} &
    \scriptsize{$48$} &
    $\frac{11}{3}$ &
    $\frac{1}{6}$ &
    -&
    -&
    -&
    -&
    -&
    -\\
&  &  &  &  &  &  &  &  &  &
\\
\hline &  &  &  &  &  &  &  &  &  &
\\
$a_{20\ell}$&
   $\frac{325}{2}$ &
    $\frac{25}{2}$ &
    - &
    - &
    -&
    -&
    -&
    -&
    -&
    -\\
&  &  &  &  &  &  &  &  &  &
\\
  \hline \hline\end{tabular}
  \end{center}
    \centerline{\parbox{10cm}{\caption{Coefficients $a_{\mu\nu\ell}$
    of the operators $T_{\mu,\nu}$}
    \label{tab1}}}
    \end{table}

\begin{table}[p]
    \vspace{.5cm}
\begin{center}
\begin{tabular}{||c||c|c|c|c|c||}
\hline \hline
 $\ell$ & 1 & 2 & 3 & 4 & 5
\\
\hline\hline
& & & & & \\
$b_{00\ell}$&
    $\frac{79}{11468800}$&
    $\frac{79}{2867200}$&
    $\frac{533}{8601600}$&
    $\frac{947}{8601600}$&
    $\frac{53}{983040}$
\\
& & & & & \\
\hline & & & & & \\
$b_{01\ell}$&
    $\frac{601}{532224000}$&
    $\frac{4757}{1596672000}$&
    $\frac{5443}{798336000}$&
    $\frac{937}{266112000}$&
    -\\
& & & & & \\
\hline & & & & & \\
$b_{10\ell}$&
    $\frac{211}{18923520}$&
    $-\frac{533}{63866880}$&
    $\frac{9127}{510935040}$&
    -&
    -\\
& & & & & \\
\hline & & & & & \\
$b_{11\ell}$&
    $\frac{1}{70963200}$&
    $\frac{1}{383201280}$&
    -&
    -&
    -\\
& & & & & \\
\hline & & & & & \\
$b_{02\ell}$&
    $-\frac{13}{479001600}$&
    $\frac{15}{958003200}$&
    -&
    -&
    -\\
& & & & & \\
\hline
& & & & & \\
$b_{20\ell}$&
    $-\frac{1}{76640256}$&
    -&
    -&
    -&
-\\
& & & & & \\
\hline \hline\end{tabular}
  \end{center}
    \centerline{\parbox{13cm}{\caption{Coefficients
    $b_{\mu\nu\ell}$
    of the matrix elements $\br x|T_{\mu,\nu}|y\kt$}
    \label{tab2}}}
    \end{table}

\ed
\begin{thebibliography}{99}
\bibitem{p2p3} A.~Mostafazadeh, J.\ Math.\ Phys.\ {\bf 43}, 2814
(2002); ibidem {\bf 43}, 3944 (2002)
\bibitem{p1} A.\ Mostafazadeh, J.\ Math.\ Phys.\ {\bf 43}, 205 (2002)
\bibitem{simon-reed} M.~Reed and B.~Simon, {\em Functional
Analysis}, vol.~1, Academic Press, San Diego, 1980
\bibitem{critique} A.~Mostafazadeh, Preprint: quant-ph/0310164
\bibitem{jpa04b} A.~Mostafazadeh and A.~Batal,
J.~Phys.~A: Math.\ Gen.\ {\bf 37}, 11645 (2004)
\bibitem{jpa05b} A.~Mostafazadeh,
J.~Phys.~A: Math.\ Gen.\ {\bf 38}, 6557 (2005)
\bibitem{p66} A.~Mostafazadeh, J.~Math.~Phys.\ {\bf 46}, 102108 (2005)
\bibitem{p4} A.~Mostafazadeh, Nucl.\ Phys.\ B, {\bf 640}, 419 (2002)
\bibitem{jmp2003} A.~Mostafazadeh, J.\ Math.\ Phys.\ {\bf 44}, 974
(2003)
\bibitem{geyer} H.~B.~Geyer, F.~G.~Scholtz, and I.~Snyman,
Czech J.~Phys.\ {\bf 54}, 1069 (2004)
\bibitem{jones} H.~F.~Jones, J.~Phys.~A: Math.\ Gen.\
{\bf 38}, 1741 (2005)
\bibitem{bbj} C.~M.~Bender, D.~C.~Brody and H.~F.~Jones, Phys.\ Rev.\
Lett.\ {\bf 89}, 270401 (2002)
\bibitem{jpa05a} A.~Mostafazadeh, J.~Phys.~A: Math.\ Gen.\ {\bf 38},
3213 (2005)
\bibitem{bbj-prd} C.~M.~Bender, D.~C.~Brody and H.~F.~Jones, Phys.\ Rev.\
D {\bf 70}, 025001 (2004)
\bibitem{wave-eqn} P.~V.~O'Neil, {\em Beginning Partial
Differential Equations}, Wiley-Intescience, New York, 1999.
\bibitem{bender-prl} C.~M.~Bender and S.~Boettcher,
Phys.\ Rev.\ Lett.\ {\bf 80}, 5243 (1998)
\bibitem{dorey} P.~Dorey, C.~Dunning, and R.~Tateo, J.~Phys.~A: Math.\ Gen.\ {\bf
34}, 5679 (2001)
\bibitem{shin} K.~C.~Shin, Commun.\ Math.\ Phys.\ {\bf 229}, 543 (2002)
\bibitem{cjp-04c} A.~Mostafazadeh, Czech J.\ Phys.\ {\bf 54}, 1125
(2004)
\bibitem{jpa-2003} A.~Mostafazadeh, J.~Phys.~A: Math.\ Gen.\
{\bf 36}, 7081 (2003)
\bibitem{tegmark} M.~Tegmark, Class.\ Quantum Grav.\ {\bf 14}, L69
(1997)
\bibitem{geyer2} F.~G.~Scholtz, H.~B.~Geyer,
and F.~J.~W.~Hahne, Ann.\ Phys.\ (NY) {\bf 213} 74 (1992)
\bibitem{ashtekar} A.~Ashtekar, J.~Lewandowski, D.~Marolf,
J.~Mour\~ao, and T.~Thiemann, J.~Math.\ Phys.\ {\bf 36}, 6456
(1995)
\end{thebibliography}
